\documentclass[print,onecolumn,showpaces,aps,showpacs]{revtex4}%
\usepackage{amsfonts}
\usepackage{amsmath}
\usepackage{amssymb}
\usepackage{graphicx}%
\providecommand{\U}[1]{\protect\rule{.1in}{.1in}}

\begin{document}

\title{Optical solitons in media with focusing and defocusing saturable
nonlinearity and a parity-time-symmetric external potential}
\author{Pengfei Li$^{1}$}

\author{Dumitru Mihalache$^{2}$}
\author{Boris A. Malomed$^{3,4}$}

\begin{abstract}
We report results for solitons in models of waveguides with focusing or
defocusing saturable nonlinearity and a parity-time ($\mathcal{PT}$%
)-symmetric complex-valued external potential of the Scarf-II type. The
model applies to the nonlinear wave propagation in graded-index optical
waveguides with balanced gain and loss. We find both fundamental and
multipole solitons for both focusing and defocusing signs of the saturable
nonlinearity in such $\mathcal{PT}$-symmetric waveguides. The dependence of
the propagation constant on the soliton's power is presented for different
strengths of the nonlinearity saturation, $S$. The stability of fundamental,
dipole, tripole, and quadrupole solitons is investigated by means of the
linear-stability analysis and direct numerical simulations of the corresponding
(1+1)-dimensional nonlinear Schr\"{o}dinger-type equation. The results show
that the instability of the stationary solutions can be mitigated or
completely suppressed, increasing the value of $S$.
\end{abstract}

\address{$^{1}$Department of Physics, Taiyuan Normal University, Taiyuan, 030031, China\\
$^{2}$Horia Hulubei National Institute of Physics and Nuclear Engineering, Magurele, Bucharest, RO-077125, Romania\\
$^{3}$Department of Physical Electronics, School of Electrical Engineering, Faculty of Engineering, Tel Aviv University, Tel Aviv 69978, Israel\\
$^4$ ITMO University, St. Petersburg 197101, Russia}

\maketitle

\section{Introduction}
Dissipative systems are usually
governed by nonlinear equations, in which the input of energy
supports self-organization, in the form of spontaneously emerging
stable  non-equilibrium structures \cite{Prigogine3,Prigogine4}.
In particular, on the basis of nonlinear dynamics and Prigogine's
idea of self-organization, dissipative solitons were studied as a
fundamental extension of solitons in conservative media \cite{Akhmediev1,Akhmediev2}.
In this context, it is relevant to stress that $\mathcal{PT}$-symmetric
systems are dissipative ones, as they include symmetrically placed gain
and loss; nevertheless, the specific symmetry makes their properties
similar to those of conservative systems, such as all-real linear spectra
and continuous families of stationary nonlinear modes.

The concept of $\mathcal{PT}$ symmetry has been drawing permanently growing
interest in the past two decades since the publication of the pioneering
paper by Bender and Boettcher \cite{1}. A simple one-dimensional $\mathcal{PT%
}$-symmetric system is described by a Schr\"{o}dinger-type Hamiltonian with
a complex-valued potential, whose real and imaginary parts must be,
respectively, even and odd functions of spatial coordinate. Although $%
\mathcal{PT}$-symmetric Hamiltonians are non-Hermitian, they may support
fully real eigenvalue spectra \cite{1,2,3}. It is worth noting that although
the introduction of $\mathcal{PT}$-symmetric models was motivated by the
quantum-mechanical setting, the concept of the $\mathcal{PT}$-symmetry has
been extended to a plethora of physical settings in photonics \cite{11,4,5},
Bose-Einstein condensates \cite{6}, plasmonic waveguides and metamaterials
in which losses appear due to metal absorption \cite{7,8,9},
superconductivity \cite{10}, and other areas.

In guided-wave optics, the propagation of light is often modeled, in the
paraxial approximation, by equations of the Schr\"{o}dinger type, with the
refractive index of a graded-index optical waveguide represented by a
real-valued effective potential. An imaginary part of the potential, if any,
stands for effects of loss and gain in the underlying waveguide. When the
gain and loss are exactly balanced, the corresponding Schr\"{o}dinger
equation includes a $\mathcal{PT}$-symmetric Hamiltonian. $\mathcal{PT}$%
-symmetric optical systems typically arise in coupled optical waveguides
with balanced gain and loss, that were firstly investigated theoretically
and experimentally in the linear regime \cite{4,11,12,13,14}. A rigorous
theoretical formulation of the $\mathcal{PT}$-symmetry in optics beyond the
paraxial approximation, based on the Maxwell's equations, was given in Ref.
\cite{15}. Furthermore, the $\mathcal{PT}$-symmetry concept was extended to
nonlinear optical systems, leading to extensive theoretical and experimental
studies. Various types of nonlinear stationary modes and their propagation
dynamics have been investigated for many types of $\mathcal{PT}$-symmetric
external potentials. In particular, bright, dark, gap, and Bragg solitons
were predicted in these models \cite%
{16,17,18,19,Radik,20,21,22,23,24,25,26,27,28,28a,Li2018}, see recent
reviews in Refs. \cite{review1,review2,review3}. It is relevant to mention
that solitons in one-dimensional $\mathcal{PT}$-symmetric periodic potentials
with saturable nonlinearity were recently considered in Refs. \cite{OC316,OC349},
where the existence of gap solitons in optical lattices and superlattices was
demonstrated. The existence of solitons in two-dimensional
$\mathcal{PT}$-symmetric periodic potentials was studied too, the results
showing that the solitons are stable at moderate powers \cite{SREP6}.
Besides that, logarithmically saturable nonlinearity can also support stable
solitons in $\mathcal{PT}$-symmetric periodic potentials \cite{EPJD70}.
In the present work, we address fundamental and higher-order families of
solitons in the one-dimensional model with saturable nonlinearity and
a $\mathcal{PT}$-symmetric potential. We demonstrate that the nonlinearity
saturation helps to suppress the instability of fundamental and
higher-order solitons.

The rest of the paper is organized as follows. In Sec. 2, the governing
model is introduced. In Sec. 3, we show that both fundamental and multipole
stationary solutions exist in the (1+1)-dimensional Schr\"{o}dinger-type
equation with both focusing and defocusing saturable nonlinearity. The
dependence of the soliton's propagation constant on the power is also
investigated. In Sec. 4, the stability of fundamental, dipole, tripole, and
quadrupole solitons is systematically investigated in the framework of the
linear-stability analysis. The evolution of perturbed stationary solitons is
studied by means of direct numerical simulations. The work is concluded by
Sec. 5.

\section{The governing model}

In the paraxial approximation, the optical wave propagation in a planar
graded-index waveguide with saturable nonlinearity is governed by the
(1+1)-dimensional nonlinear Schr\"{o}dinger equation, written here in the
scaled form:
\begin{equation}
i\frac{\partial \psi }{\partial \zeta }+\frac{\partial ^{2}\psi }{\partial
\xi ^{2}}+U\left( \xi \right) \psi +\sigma \frac{\left\vert \psi \right\vert
^{2}\psi }{1+S\left\vert \psi \right\vert ^{2}}=0,  \label{2.1}
\end{equation}%
where\ $\psi \left( \zeta ,\xi \right) $ is the envelope of the
electromagnetic field, $U\left( \xi \right) \equiv V\left( \xi \right)
+iW\left( \xi \right) $ is a complex-valued potential, and $\sigma =\pm 1$
represent focusing ($+$) and defocusing ($-$) signs of the nonlinearity,
respectively. Finally, the coefficient $S$ defines the saturation of the
nonlinearity \cite{29,30}.

Stationary solutions are sought for in the usual form, $\psi \left( \zeta
,\xi \right) =\phi \left( \xi \right) e^{i\beta \zeta }$, where $\phi \left(
\xi \right) $ is a complex function, and $\beta $ is a real propagation
constant. Substitution of this in equation (\ref{2.1}) yields
\begin{equation}
\frac{d^{2}\phi }{d\xi ^{2}}+U\left( \xi \right) \phi \left( \xi \right)
+\sigma \frac{\left\vert \phi \right\vert ^{2}\phi }{1+S\left\vert \phi
\right\vert ^{2}}-\beta \phi \left( \xi \right) =0.  \label{2.2}
\end{equation}

In this paper,\ we consider the nonlinear model with the $\mathcal{PT}$%
-symmetric Scarf-II potential \cite{Scarf,Grosche} of the form
\begin{equation}
V\left( \xi \right) =V_{0}\text{sech}^{2}\left( \frac{\xi }{\xi _{0}}\right)
,  \label{2.3}
\end{equation}%
\begin{equation}
W\left( \xi \right) =W_{0}\text{sech}\left( \frac{\xi }{\xi _{0}}\right)
\text{tanh}\left( \frac{\xi }{\xi _{0}}\right) ,  \label{2.4}
\end{equation}%
where $V_{0}$ and $W_{0}$ represent the strengths of its real and imaginary
parts, and $\xi _{0}$ defines the width of the $\mathcal{PT}$-symmetric
potential.

\begin{figure}[th]
\centering\includegraphics[width=3in]{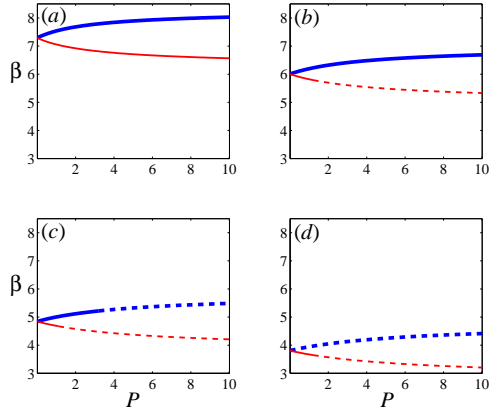}
\caption{The propagation constants of stationary solutions versus the
soliton's integral power. Thick blue solid and dashed curves correspond to
the stable and unstable stationary solutions with focusing saturable
nonlinearity, and thin red solid and dashed curves correspond to the stable
and unstable stationary solutions with defocusing saturable nonlinearities,
respectively. Panels (\textit{a}), (\textit{b}), (\textit{c}), and (\textit{d%
}) display the results for the fundamental, dipole, tripole, and quadrupole
solitons, respectively. Here, parameters in (\protect\ref{2.3}) and (\protect
\ref{2.4}) are $V_{0}=8$, $W_{0}=4$, $\protect\xi _{0}=4$, and $S=1$. In fact,
results of more extensive numerical calculations demonstrate that relation
$W_0=(1/2)V_0$, implied by these values, adequately represents the generic situation.}
\label{fig1}
\end{figure}

\begin{figure}[th]
\centering\includegraphics[width=4in]{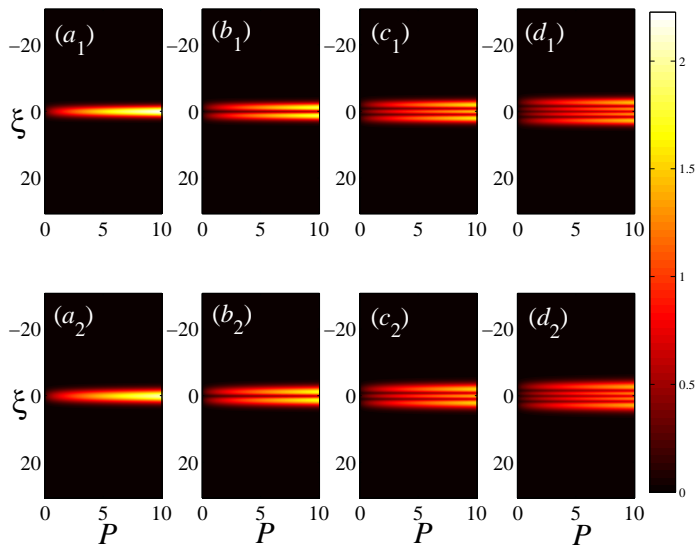}
\caption{Evolution of profiles of the stationary solutions with the increase
of the solitons' power. Panels (\textit{a}$_{1,2}$), (\textit{b}$_{1,2}$), (%
\textit{c}$_{1,2}$), and (\textit{d}$_{1,2}$) represent fundamental, dipole,
tripole, and quadrupole solitons, respectively, with subscripts 1 and 2
corresponding to the focusing and defocusing sign of the saturable
nonlinearity. Parameters are the same as in figure 1.}
\label{fig2}
\end{figure}

\section{The analysis of the stationary solutions}

We seek for solutions to equation (\ref{2.1}) with both focusing and
defocusing sign of the nonlinearity, i.e., $\sigma =+1$ and $\sigma =-1$,
respectively. As an example, which adequately represents the generic case,
we choose $V_{0}=8$, $W_{0}=4$, and $\xi _{0}=4$ in equations (\ref{2.3})
and (\ref{2.4}) .

The numerical results demonstrate that there exist both fundamental and
higher-order stationary solutions of equation (\ref{2.2}). The dependence of
the nonlinear propagation constants on the soliton's integral power, $%
P=\int\nolimits_{-\infty }^{+\infty }\left\vert \phi \right\vert ^{2}d\xi $,
is presented in figure 1, where blue and red curves correspond to the case
of focusing and defocusing saturable nonlinearities, respectively. Figure 1
demonstrates that the $\beta (P)$ curves originate, at $P=0$, from values
corresponding to the respective modes generated by the linear Schr\"{o}%
dinger equation with the potential (\ref{2.3}), (\ref{2.4}). Further, figure
1 shows too that the slope of the dependence is positive and negative, i.e.,
$d\beta /dP>0$ and $d\beta /dP<0$, respectively, for the focusing and
defocusing sign of the nonlinearity. In either case, the sign complies,
severally, with the Vakhitov-Kolokolov (VK) \cite{VK1,VK2} or anti-VK \cite%
{antiVK} criteria, which are necessary conditions for the stability of
solitons created by focusing or defocusing nonlinearity of any type,
including the saturable one. The profiles of the corresponding solitons, for
$P$ increasing from $0.01$ to $10$, are depicted in figure 2.

It is also relevant to address the dependence of the propagation constant, $%
\beta $, of fundamental solitons on the integral power, $P$, for different
values of the saturation parameter $S$. The saturation slows the variation
of $\beta $ with the growth of $P$, for both the focusing and defocusing
signs of the nonlinearity, as shown in figure 3. Naturally, when $S$ is very
small (e.g., $S=0.1$) in figure 3, the $\beta (P)$ dependence is close to
that obtained for the cubic nonlinearity.

\begin{figure}[th]
\centering\includegraphics[width=5in]{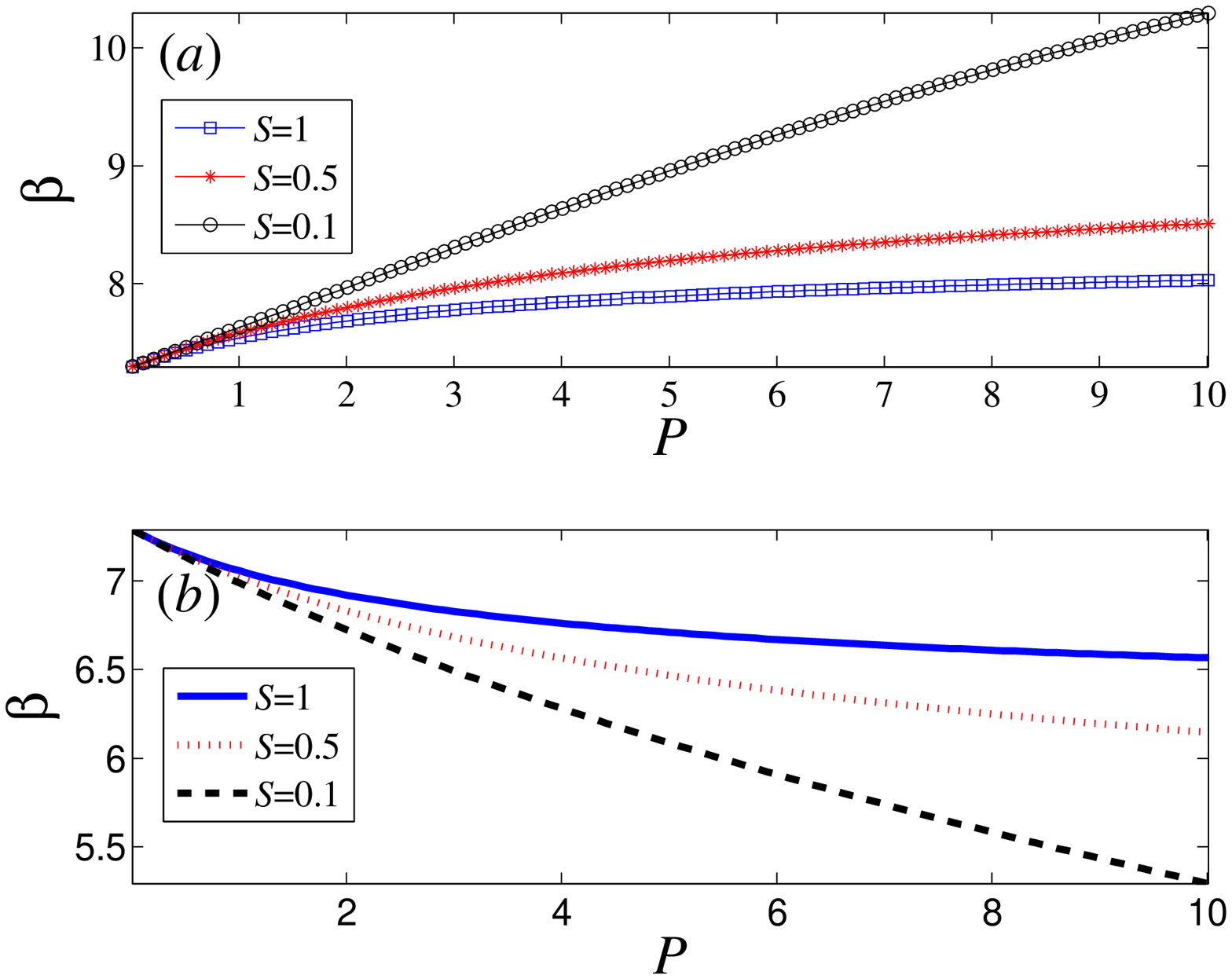}
\caption{The propagation constant of fundamental solitons versus their
integral power, with different values of saturation parameter $S$, see Eqs. (%
\protect\ref{2.3}) and (\protect\ref{2.4}), for the self-focusing (\textit{a}%
) and the defocusing (\textit{b}) sign of the nonlinearity. Other parameters
are the same as in figure 1.}
\label{fig3}
\end{figure}

\section{The linear-stability analysis and propagation dynamics of solitons}

To examine the stability of the stationary solutions $\phi (\xi )$ with
propagation constant $\beta $, we added small perturbations $u\left( \xi
\right) $ and $v\left( \xi \right) $ to them, in the usual form,
\begin{equation}
\psi \left( \zeta ,\xi \right) =e^{i\beta \zeta }\left[ \phi \left( \xi
\right) +u\left( \xi \right) e^{\delta \zeta }+v^{\ast }\left( \xi \right)
e^{\delta ^{\ast }\zeta }\right] ,  \label{3.1}
\end{equation}%
where $\delta $ is the (complex) instability growth rate sought for. The
substitution of equation (\ref{3.1}) into equation (\ref{2.1}) and
subsequent linearization leads to the eigenvalue problem for $\delta $:%
\begin{equation}
i\left(
\begin{array}{cc}
L_{11} & L_{12} \\
L_{21} & L_{22}%
\end{array}%
\right) \left(
\begin{array}{c}
u \\
v%
\end{array}%
\right) =\delta \left(
\begin{array}{c}
u \\
v%
\end{array}%
\right) ,  \label{3.2}
\end{equation}%
with operators
\begin{equation}
L_{11}=\frac{d^{2}}{d\xi ^{2}}+U\left( \xi \right) -\beta +\frac{2\sigma
\left\vert \phi \right\vert ^{2}}{1+S\left\vert \phi \right\vert ^{2}}-\frac{%
\sigma S\left\vert \phi \right\vert ^{4}}{\left( 1+S\left\vert \phi
\right\vert ^{2}\right) ^{2}},  \label{3.3}
\end{equation}%
\begin{equation}
L_{12}=\frac{2\sigma \phi ^{2}}{1+S\left\vert \phi \right\vert ^{2}}-\frac{%
\sigma S\left\vert \phi \right\vert ^{2}\phi ^{2}}{\left( 1+S\left\vert \phi
\right\vert ^{2}\right) ^{2}},  \label{3.4}
\end{equation}%
\begin{equation}
L_{21}=\frac{\sigma S\left\vert \phi \right\vert ^{2}\phi ^{\ast 2}}{\left(
1+S\left\vert \phi \right\vert ^{2}\right) ^{2}}-\frac{2\sigma \phi ^{\ast 2}%
}{1+S\left\vert \phi \right\vert ^{2}},  \label{3.5}
\end{equation}%
\begin{equation}
L_{22}=-\left[ \frac{d^{2}}{d\xi ^{2}}+U^{\ast }\left( \xi \right) -\beta +%
\frac{2\sigma \left\vert \phi \right\vert ^{2}}{1+S\left\vert \phi
\right\vert ^{2}}-\frac{\sigma S\left\vert \phi \right\vert ^{4}}{\left(
1+S\left\vert \phi \right\vert ^{2}\right) ^{2}}\right] .  \label{3.6}
\end{equation}%
The stationary solution $\phi \left( \xi \right) $\ is linearly unstable if
there is at least one eigenvalue with Re$(\delta )>0$.

\begin{figure}[th]
\centering\includegraphics[width=4.5in]{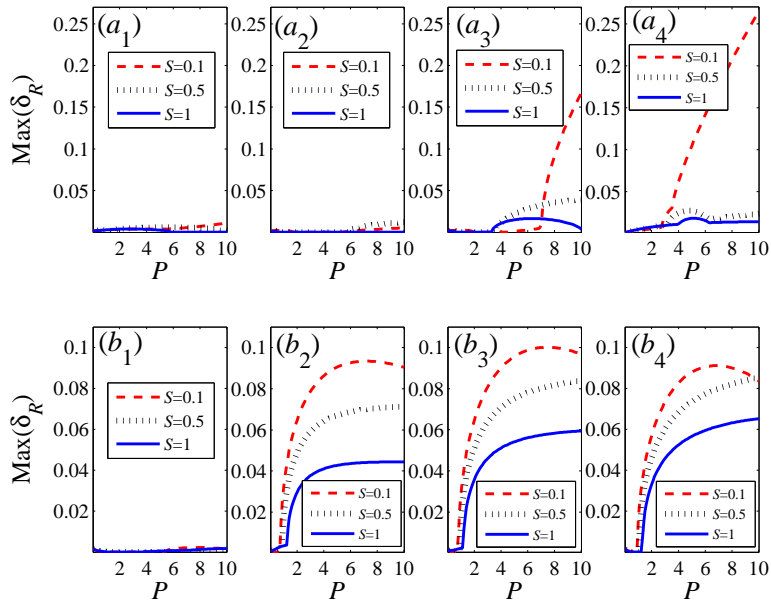}
\caption{Eigenvalues of the linear-stability analysis, produced by the
numerical solution of equation (\protect\ref{3.2}) for different values of
saturation parameter $S$, see equation (\protect\ref{2.1}). Panels (\textit{a%
}$_{1}$), (\textit{a}$_{2}$), (\textit{a}$_{3}$), and (\textit{a}$_{4}$)
depict the largest real part of $\protect\delta $ for the fundamental,
dipole, tripole, and quadrupole soliton, respectively, in the case of\ the
focusing saturable nonlinearity. Panels (\textit{b}$_{1}$), (\textit{b}$_{2}
$), (\textit{b}$_{3}$), and (\textit{b}$_{4}$) show the same for the
defocusing nonlinearity. Parameters are the same as in figure 1.}
\label{fig4}
\end{figure}

Figure 4 presents the dependence of the largest instability growth rate, $%
\mathrm{Max}\left( \delta _{R}\right) $, on the solitons' integral power, $P$%
, as produced by numerical solution of the eigenvalue problem for the
fundamental, dipole, tripole, and quadrupole solitons at different values of
$S$ and both signs of the nonlinearity, focusing and defocusing (the top and
bottom rows in figure 4, respectively). The numerical results displayed in
panels 4(\textit{a}$_{1}$) and 4(\textit{a}$_{2}$) show that, under the
action of the self-focusing nonlinearity, Max($\delta _{R}$) is negligibly
small for the fundamental and dipole solitons, which indicates that they are
linearly stable, irrespective of the value of $S$. On the other hand, the
tripoles and quadrupoles are stable when the solitons' power is sufficiently
small, as seen in panels 4(\textit{a}$_{3}$) and 4(\textit{a}$_{4}$) (in
fact, the quadrupoles are almost completely unstable). Further, under the
action of self-defocusing nonlinearity, fundamental solitons are linearly
stable, irrespective of $S$, as shown in figure 4(\textit{b}$_{1}$), while
panels 4(\textit{b}$_{2}$), 4(\textit{b}$_{3}$), and 4(\textit{b}$_{4}$)
show that the dipoles, tripoles, and quadrupoles are stable, in this case,
only in tiny intervals of very small powers. A general trend, suggested by
these results, is that the instability is attenuated with the increase of
the saturation strength $S$.

\begin{figure}[th]
\centering\includegraphics[width=4.5in]{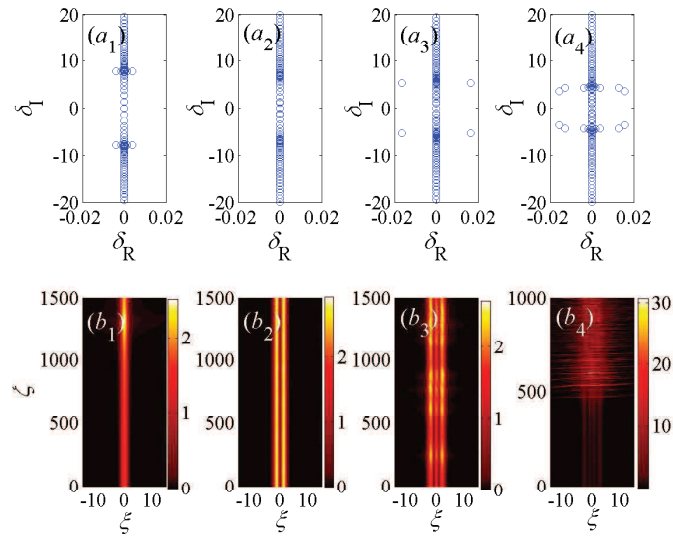}
\caption{Eigenvalue spectra produced by the linear-stability analysis for
different soliton species, and the corresponding evolution plots for the
focusing nonlinearity ($\protect\sigma =+1$) and $S=1$. Panels (\textit{a}$%
_{1}$,\textit{b}$_{1}$), (\textit{a}$_{2}$,\textit{b}$_{2}$), (\textit{a}$%
_{3}$,\textit{b}$_{3}$), and (\textit{a}$_{4}$,\textit{b}$_{4}$) pertain,
respectively, to a fundamental soliton with power $P=5$, a dipole with $P=10$%
, a tripole with $P=6$, and a quadrupole with $P=6$. Other parameters are
the same as in figure 1.}
\label{fig5}
\end{figure}

\begin{figure}[th]
\centering\includegraphics[width=4.5in]{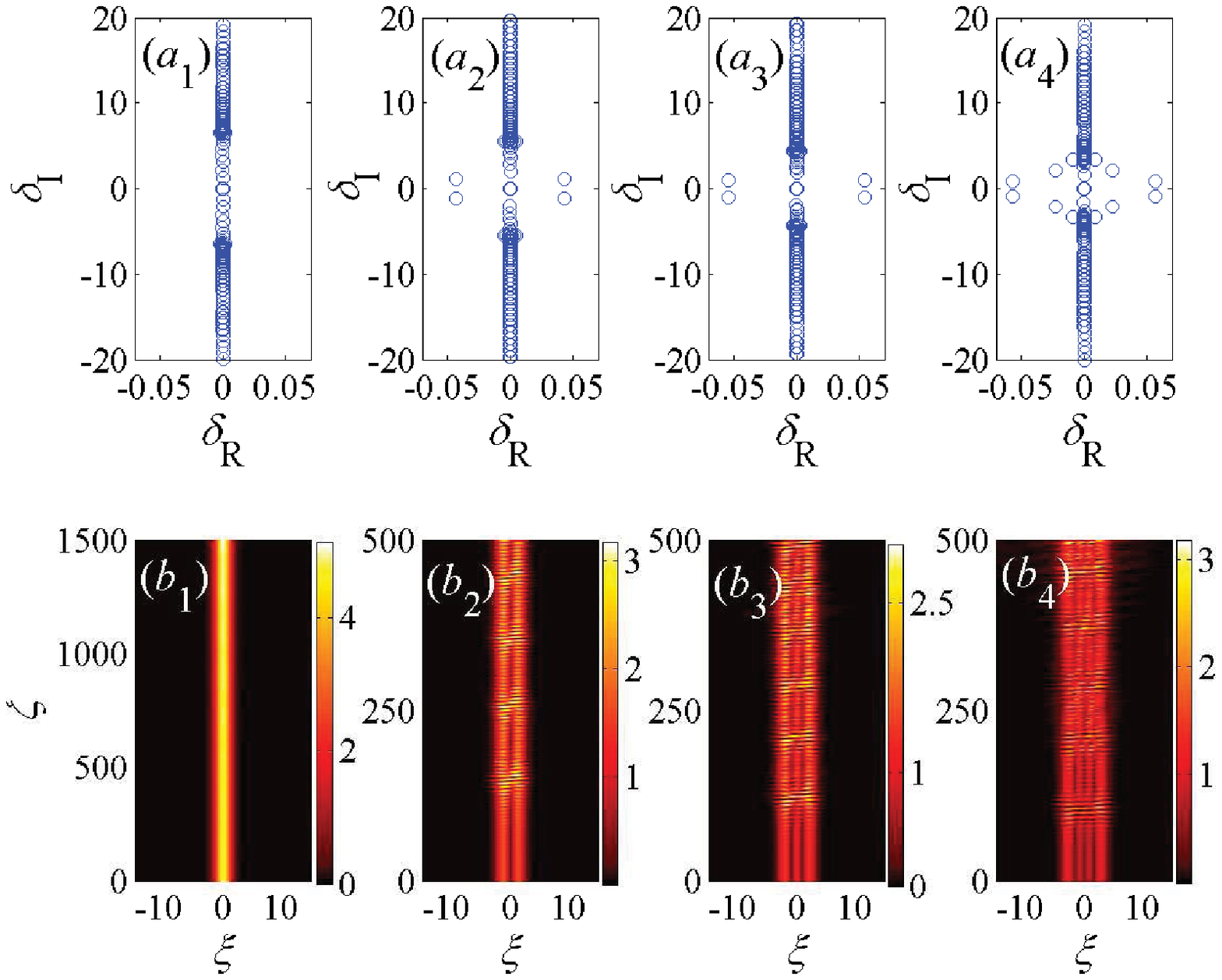}
\caption{The same as in figure 5, but in the case of the defocusing
nonlinearity ($\protect\sigma =-1$). Panels (\textit{a}$_{1}$,\textit{b}$%
_{1} $), (\textit{a}$_{2}$,\textit{b}$_{2}$), (\textit{a}$_{3}$,\textit{b}$%
_{3}$), and (\textit{a}$_{4}$,\textit{b}$_{4}$) pertain, respectively, to a
fundamental soliton with power $P=10$, and dipole, tripole, and quadrupole
solitons, all with $P=6$.}
\label{fig6}
\end{figure}

The results of the linear-stability analysis are confirmed, separately in
figures 5 and 6 for the focusing and defocusing nonlinearities, by direct
numerical simulations of the perturbed evolution of solitons, performed in
the framework of equation (\ref{2.1}). In these figures, the results of the
simulations are displayed along with (in)stability spectra of the underlying
stationary states. The eigenvalue spectra of equation (\ref{3.2}) for
fundamental, dipole, tripole, and quadrupole solitons are shown on the top
row in figure 5. In particular, figures 5(\textit{b}$_{1}$) and 5(\textit{b}$%
_{2}$) confirm the stable propagation of the fundamental and dipole solitons
in the case of self-focusing nonlinearity, under the action of sufficiently
strong random perturbations, with relative amplitudes $10\%$. Very small
nonzero values of Re$(\delta )$, which are seen, for these solutions, in
Figs. 5(\textit{a}$_{1}$) and 5(\textit{a}$_{2}$), but do not lead to any
instability in the direct propagation, are a manifestation of a small
inaccuracy of the numerical analysis. On the other hand, the tripoles and
quadrupoles are definitely unstable, in accordance with the prediction of
the linear-stability analysis, as seen in figures 5(\textit{b}$_{3}$) and 5(%
\textit{b}$_{4}$). It is observed that the unstable tripole is spontaneously
transformed into a breather, while the quadrupole is, eventually, destroyed
by the instability.

Similar results for the defocusing nonlinearity are summarized in figure 6,
where, in accordance with the prediction of the linear-stability analysis,
the fundamental soliton is stable, while dipole, tripole, and quadrupole are
not, transforming into breathers.

\section{Conclusion}

We have performed a systematic study of solitons in the models of waveguides
with the self-focusing and defocusing saturable nonlinearity and $\mathcal{PT%
}$-symmetric complex-valued potential of the Scarf-II type. Both models
support families of fundamental, dipole, tripole, and quadrupole solitons.
Their stability has been investigated by means of the linear-stability
analysis, and confirmed by direct numerical simulations. In particular,
fundamental solitons are completely stable for both signs of the
nonlinearity and, quite interestingly, dipoles and a part of the tripole
family are stable too in the case of self-focusing nonlinearity. Other
soliton families are practically completely unstable, including all
higher-order solitons under the action of self-defocusing nonlinearity. A
general trend is that the increase of the saturation strength partly
suppresses the instability.

As an extension of the analysis, it may be interesting to perform it for
more general shapes of the $\mathcal{PT}$-symmetric external potential.

\section{Acknowledgement}
This work was supported by Doctoral Scientific Research Foundation
of Taiyuan Normal University No. I170144, and by the Israel Foundation through grant No. 1287/17.


\end{document}